\documentstyle[epsf]{mn}

\title[Negative superhumps in V1405 Aql]
{Detection of negative superhumps in a LMXRB -- an end to the long
debate on the nature of V1405~Aql (X1916-053)}

\author[A. Retter et al.]
   {A.~Retter,$^{1}$\thanks {Email: retter@physics.usyd.edu.au}$^{,2}$
Y. Chou,$^3$ T. R. Bedding$^{1}$ and T. Naylor$^{4,2}$\\
 $^1$School of Physics, University of Sydney, 2006, Australia\\
 $^2$Dept. of Physics, Keele University, Keele, Staffordshire, ST5 5BG\\
 $^3$Dept. of Physics, National Tsing Hua University, No. 101, Sec. 2, Kuang 
Fu Rd, Hsinchu, Taiwan\\
 $^4$School of Physics, University of Exeter, Stocker Road, Exeter, EX4 4QL\\}

\date{Accepted 2001 December 18. Received 2001 December 5; in original form 
2001 July 27}

\begin{document}

\maketitle

\begin{abstract}

The detection of two similar periodicities (3001 and 3028~s) in the light 
curve of V1405~Aql, a low mass X-ray Binary (LMXRB), has attracted the 
attention of many observers. Two basic competing models have been offered 
for this system. According to the first, V1405~Aql is a triple system. The 
second model invokes the presence of an accretion disc that precesses in 
the apsidal plane, suggesting that the shorter period is the orbital period 
while the longer is a positive superhump. The debate on the nature of 
V1405~Aql has been continued until very recently. Re-examination of 
previously published X-ray data reveals an additional periodicity of 2979~s, 
which is naturally interpreted as a negative superhump. The recently found 
4.8-d period is consequently understood as the nodal precession of the disc. 
This is the first firm detection of negative superhumps and nodal precession 
in a LMXRB. Our results thus confirm the classification of V1405~Aql as a 
permanent superhump system. The 14-year argument on the nature of this 
intriguing object has thus finally come to an end. We find that the ratio 
between the negative superhump deficit (over the orbital period) and the 
positive superhump excess is a function of orbital period in systems that 
show both types of superhumps. This relation presents some challenge to 
theory as it fits binaries with different components. We propose that a 
thickening in the disc rim, which causes increased occultation of the X-ray 
source, is the mechanism responsible for both types of superhumps in LMXRBs. 
However, the positive signal is related only to the pronounced dips in the 
light curve, where the point-like central source is covered up, whereas the 
morphology of the negative superhump signal appears quite smooth, implying 
obscuration of a larger X-ray emitting region, possibly the inner accretion 
disc or a corona. According to our model superhumps (both in the X-ray and 
optical regimes) are permitted in high inclination LMXRBs contrary to 
Haswell et al. (2001) prediction.

\end{abstract}

\begin{keywords}
accretion, accretion discs -- X-rays: binaries -- 
X-rays: individual: V1405~Aql -- novae, cataclysmic variables 

\end{keywords}

\section{Introduction}
\subsection{Permanent superhumps}

Permanent superhumps have been observed so far in about 20 cataclysmic 
variables (CVs) (Patterson 1999). These show superhumps (quasi-periodicities 
shifted by a few percent from their orbital periods) in their optical light 
curves during normal brightness state. In contrast, SU~UMa systems (see 
Warner 1995 for a review of SU~UMa systems and CVs in general) have 
superhumps only during their bright dwarf nova outbursts (superoutbursts). 

Permanent superhumps can either be a few percent longer than the orbital 
periods and they are called `$\bf positive$ $\bf superhumps$', or shorter
-- `$\bf negative$ $\bf superhumps$'. The positive
superhump is explained as the beat between the binary motion and the 
precession of an accretion disc in the apsidal plane. Similarly, the 
negative superhump is understood as the beat between the orbital period and 
the nodal precession of the disc (Patterson 1999).

According to theory, superhumps are developed when the accretion disc 
extends beyond the 3:1 resonance radius and becomes elliptical (Whitehurst 
\& King 1991). Osaki (1996) suggested that permanent superhump systems 
differ from other subclasses of non-magnetic CVs in their relatively short 
orbital periods and high mass-transfer rates, resulting in accretion discs 
that are thermally stable but 
tidally unstable. Retter \& Naylor (2000) gave observational support to 
this idea.  Simulations suggest that superhumps can only occur in binary 
systems with small mass ratios -- $q=M_{donor}/M_{compact}\la0.33$ 
(Whitehurst 1988; Whitehurst \& King 1991; Murray 2000). Although this 
condition is easily met in short orbital period LMXRBs, whose primaries 
are compact massive objects, superhumps have been only seen in a few LMXRBs 
in outburst (e.g. O'Donoghue \& Charles 1996). Here we report the detection 
of permanent superhumps in a persistent LMXRB.

\subsection{V1405~Aql (X1916-053 / 4U 1915-05)}

V1405~Aql was first discovered as an X-ray source showing type-I bursts 
(Becker et al. 1977; Doxsey et al. 1977) suggesting that its primary star is 
a neutron star. White \& Swank (1982) and Walter et al. (1982) independently 
found $\sim$3000-s periodic dips in its X-ray light curve. The dips are 
believed to occur due to obscuration of the disc edge (probably where the 
accretion stream impacts the disc). Such a short orbital period implies a 
low mass hydrogen-deficient secondary star (Nelson, Rappaport \& Joss 1986). 
The 21-mag optical counterpart of V1405~Aql was identified by Schmidtke 
(1988) and Grindlay et al. (1988). They also reported a detection of an 
optical periodicity, about one percent longer than the X-ray period. The 
difference between the X-ray and optical periods was confirmed by further 
extensive observations of V1405~Aql (Smale et al. 1988; 1989; 1992; Callanan 
1993; Callanan, Grindlay \& Cool 1995; Yoshida et al. 1995; Church et al. 
1997; 1998; Ko et al. 1999; Morley et al. 1999; Chou, Grindlay \& Bloser 
2001; Homer et al. 2001). 

The various models offered so far for the periodicities found in V1405~Aql 
can be grouped into two basic models -- a triple system (Grindlay 1986; 
Grindlay et al. 1988) and superhumps (Schmidtke 1988; White 1989). 
According to the first model, which was motivated by the detection of a 
possible 199-d periodicity in the X-ray light curve of V1405~Aql 
(Priedhorsky \& Terrell 1984), the longer 3028-s period is the binary inner 
orbital period, while the shorter 3001-s period is the beat between 
the binary period and the $\sim$4-d orbital period of a third companion. 
The 199-d period is explained by the eccentricity of the inner binary 
orbit (Chou et al. 2001). The second model suggests that the 3001-s period 
is the binary period and that the 3028-s period is a positive superhump. 
We note that both models predict the presence of a beat periodicity of 
$\sim$4 d. Indeed evidence for a period of $\sim$3.9~d has been found in the 
X-ray light curve of V1405~Aql (Chou et al. 2001; see also Homer et al. 2001).

The debate on the nature on V1405~Aql has continued (Chou et al. 2001; 
Homer et al. 2001; Haswell et al. 2001), although some preference was
given to the superhump scenario. One of the arguments against the triple 
system model is that the 199-d period was not confirmed by further 
observations, although a second possible evidence for this periodicity was 
found by Smale \& Lochner (1992). On the other hand, an argument against 
the superhump model is that the 3028-s period is quite stable while 
superhump periods are unstable (Callanan et al. 1993; 1995; Chou et al. 
2001; Homer et al. 2001).

There is a strong observational link between positive and negative 
superhumps in CVs as light curves of several systems show both (Patterson 
1999; Arenas et al. 2000; Retter et al. 2002b). In addition, Patterson (1999) 
found that period deficits in negative superhumps are about half period 
excesses in positive superhumps: $\epsilon_{-}\approx-0.5\epsilon_{+}$, 
where $\epsilon=(P_{superhump}-P_{orbital})/P_{orbital}$. Thus, we decided 
to look in available photometric data on V1405~Aql for negative superhumps, 
which would be predicted to have a period near 2986~s. Indeed we have found 
a new periodicity in the X-ray light curve of V1405~Aql. These findings were 
reported by Retter, Chou \& Bedding (2002a), and are presented below. We also 
present a new relation for systems that have both types of superhumps, and 
discuss the mechanisms for superhumps in the X-ray and optical regimes.

\section{Observations and Analysis}

We have re-analysed existing X-ray photometry that was presented by Chou 
et al. (2001). The dataset consists of about 150 ksec of observations by 
the RXTE satellite obtained during 17 occasions between 1996 February to 
October.

In Figs.~1a $\&$ 2a the power spectrum (Scargle 1982) of 10 successive 
runs in 1996 May is presented. No de-trending method was used. Bursts were 
not observed during these observations. In addition to the two known periods 
(3001 $\&$ 3028~s, marked as f$_{1}$ $\&$ f$_{2}$) and their 1-d$^{-1}$ 
aliases, there is a third peak (labelled f$_{3}$) together with its 
1-d$^{-1}$ alias pattern. After fitting and subtracting the first term of 
the two known frequencies (f$_{1}$ $\&$ f$_{2}$), the third, which 
corresponds to the periodicity 0.034483$\pm$0.000013 d (2979.3$\pm$1.1 s), 
becomes the strongest peak in the residual power spectrum (Figs.~1b $\&$ 2b). 
The error was calculated by 1000 simulations (see e.g. Retter et al. 2002b). 
The subtraction of nine more terms of the periods (to account for a 
non-sinusoidal shape) and the 3.9-d period (whose peak in the raw power 
spectrum is not significant) yields similar results. This point is valid for 
all tests listed below. The power spectrum of the whole 1996 dataset 
resembles the result of the 10 runs in May, however, it is noisier as the 
gaps between the observations are long. 

\begin{figure}

\centerline{\epsfxsize=3.0in\epsfbox{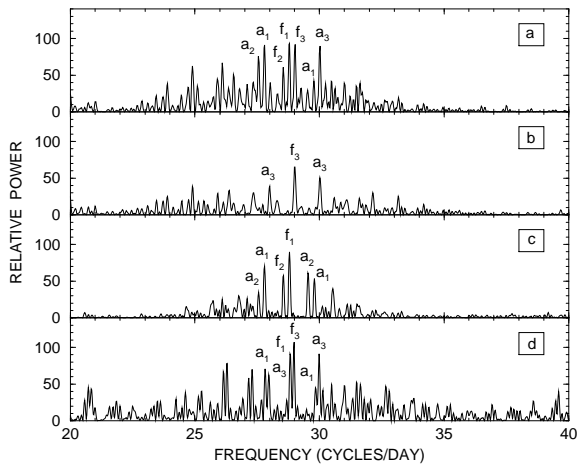}}

\caption{Power spectra of 10 successive runs in 1996 May: 
{\bf a}. Raw data; In addition to the two previously known periods -- the 
3001-s period (marked as f$_{1}$) and the 3028-s period (f$_{2}$), there is 
a third structure of peaks centered around 2979~s (f$_{3}$); `a$_{i}$' 
(i=1-3) represent 1-d$^{-1}$ aliases of `f$_{i}$'.
{\bf b}. After fitting and subtracting f$_{1}$ and f$_{2}$, f$_{3}$ becomes 
the strongest peak in the power spectrum.
{\bf c}. Power spectrum of a synthetic light curve, consisting of sinusoids 
of the two known periods (plus noise) sampled as the data thus illustrating 
the window function. 
{\bf d}. Same as (a), after rejecting the dips.} 

\end{figure}

\begin{figure}

\centerline{\epsfxsize=3.0in\epsfbox{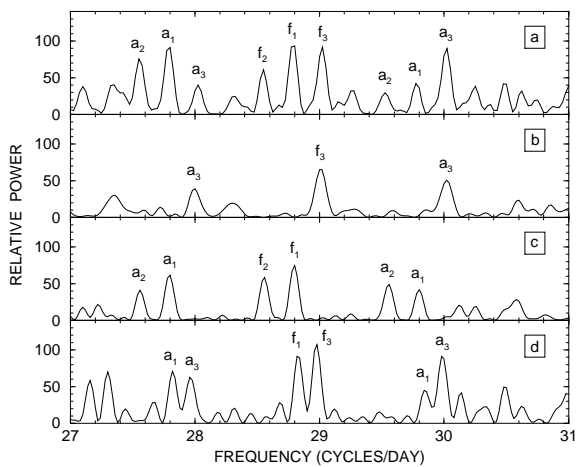}}

\caption{Same as Fig.~1, zoomed into the periods.}

\end{figure}

In Fig.~3 we show the mean shape of the 2979-s period in 1996 May. The 
full amplitude of the sinusoidal fit to the light curve is 8.0$\pm$1.2 
counts/sec, which translates to 25$\pm$4\% of the mean count rate. This 
is smaller than the mean amplitude of the 3028-s period -- 28$\pm$4\% 
and the mean depth of the 3001-s dips -- 35$\pm$5\%. The superposition
of these three signals can explain the deepest observed dips that 
sometimes reach almost 100\% (i.e. total obscuration).

\begin{figure}

\centerline{\epsfxsize=3.0in\epsfbox{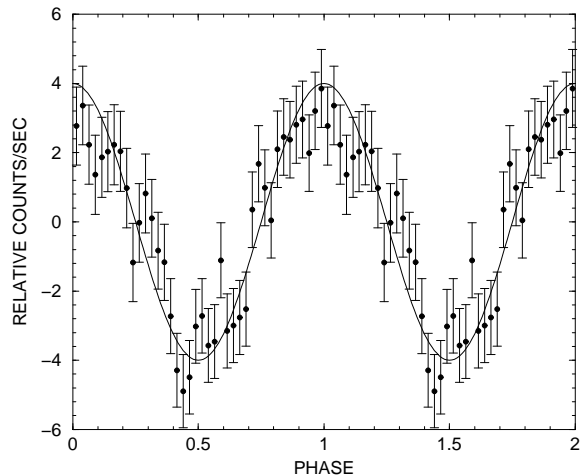}}

\caption{The data folded on the 2979-s period and binned into 
40 bins. The solid line represents the sinusoidal fit to the data.}

\end{figure}

\subsection{Tests}

\subsubsection{The window function}

To check whether the third peak in the power spectrum could be an artifact 
of the window function, a noiseless simulation of the 1996 May set was 
created. A synthetic light curve was built using two sinusoids at the 
3001 and 3028-s periods. These sinusoids were given the same amplitudes 
they have in the data and sampled according to the window function. There 
was no evidence for significant power at the proposed period. 

\subsubsection{Uncorrelated noise}

In an attempt to check whether uncorrelated noise could be responsible for 
the presence of the new peak in the power spectrum, the two previously 
known periods (3001 and 3028 s) were subtracted from the data. The remaining 
points, assumed to be white noise, were shuffled randomly between the 
timings. A power spectrum was calculated for each random configuration. The 
highest peak at the range f=20-40 d$^{-1}$ in 1000 simulations did not reach 
the height of f$_{3}$.

In a different test, that checks the significance of a third peak in the 
presence of two others, we took the synthetic light curves (see previous 
section) and now added noise, defined as the root mean square of the 
original data minus the periods modelled. We then searched for the highest 
peak in a small interval (28.9-29.1 cycles/day) around the candidate period 
(f$_{3}$). The lower value is dictated by the nearby presence of the 3001-s 
period (f$_{1}$). In 1000 simulations, no peak reached the height of the 
candidate periodicity. Figs.~1c $\&$ 2c show an individual example of this 
simulation. 

\subsubsection{Correlated noise}

Another test we tried was to assess the probability that correlated 
noise could be responsible for the candidate periodicity. In the absence 
of a model for the correlated noise, the best test is to use the 
repeatability between different datasets. Thus the 1996 May data were
divided into two subsets -- the first and last five runs. Power spectra
of both sets were calculated after the two known periods (3001 $\&$ 3028~s)
were removed. Fig.~4 displays the results. In both sets the strongest peak 
(up to a 1-d$^{-1}$ alias) is very near the peak at the power spectrum of the 
whole set (Figs.~1 $\&$ 2). 

\begin{figure}

\centerline{\epsfxsize=3.0in\epsfbox{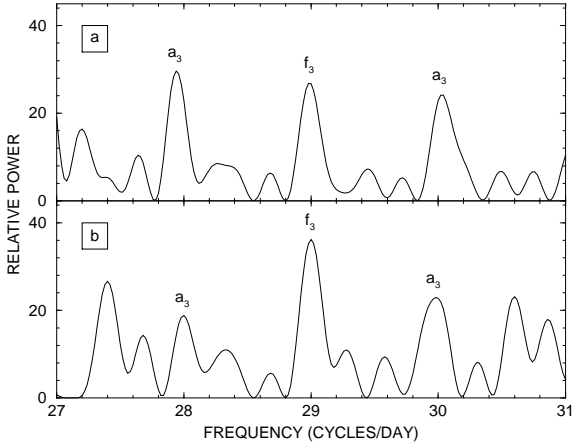}}

\caption{Power spectra of two parts of the 10 successive runs in 1996 May
after the 3001 and 3028-s periods have been removed:
{\bf a}. First five runs.
{\bf b}. Last five runs.
In both panels the highest peak or a 1-d$^{-1}$ alias corresponds to the same 
periodicity.}

\end{figure}

Given that we have found a period in the first subset, we can ask how 
likely it is that the strongest period in the second subset would be 
consistent with it. The probability of the highest peak in another dataset 
being, by chance, compatible with the candidate period in the first set is 
0.04. This was calculated from (i) twice the frequency error (0.02 d$^{-1}$) 
on the peak in the first subset and (ii) the range over which it could occur 
taken as the spacing of the 1-d$^{-1}$ aliases. This test suggests that the 
candidate periodicity is 96\% significant. We note, however, that the 
frequency error was calculated by 1000 simulations. A naive estimate of the 
error as the width at half maximum of the peak yields a somewhat lower 
significance level.

\subsubsection{The influence of the dips}

To eliminate the possibility that the new periodicity in 
V1405~Aql arises from random variations in the shape, width and / 
or depth of the dips, all points connected with the dips were manually 
rejected. The highest peak in the power spectrum of the remaining data is 
consistent with f$_{3}$ (Figs.~1d $\&$ 2d). This test shows that the 
`persistent' light curve (i.e. without the dips) is dominated by the 2979-s 
periodicity.

\section{Discussion}

\subsection{The new period}

A period of 2979~s has been found in the X-ray light curve of V1405~Aql 
in addition to the other known periods. This feature was actually 
noticed by Chou et al. (2001) who mistakenly identified it with a 3.9-d 
sideband of the 3001-s period. It is, however, $\sim5\sigma$ away from 
the expected sideband at 2974.2 s and we have shown above that it is not 
an alias. We also note that a period of 2984.6$\pm$6.8 s was detected in 
the Ginga data of this object (Smale et al. 1989). As it is inconsistent 
with either the 3001 and 3028-s periods, it probably represents the new 
periodicity as well. 

The 2979-s period is shorter than the 3001-s period by about 0.7 percent. 
Assuming that the 3001-s period is the orbital period and that the 
3028-s period is a positive superhump, the new period is naturally 
explained as a negative superhump. Our suggestion implies a nodal 
precession of $\sim$4.8 d. Indeed Chou et al. (2001) found that the 
phase jitter of the X-ray dips in the 1996 May data is modulated with a 
period of 4.86 d (as well as with the $\sim$3.9-d period). Homer et al. 
(2001) reached a similar conclusion from a different dataset and found a 
period of 4.74$\pm$0.05 d. Thus, the superhump model can explain this
periodicity as well.

The 2979-s period in the X-ray light curve of V1405~Aql and the beat 
period at $\sim$4.8 d cannot be explained by the triple system model. 
This scenario was motivated by the possible detection of the 199-d period 
that has not been confirmed by further observations (Section 1.2). Homer et 
al. (2001) suggested a few other possible periodicities ($\sim$8.6, $\sim$12, 
$\sim$23, $\sim$83 d) in the X-ray light curve of V1405~Aql. It is very 
likely that V1405~Aql has random variations of the order of a few tens of 
day, but not a coherent long-term periodicity. We thus believe that the 
triple system model for V1405~Aql is finally ruled out.

Table~1 lists the periods observed in V1405~Aql and their interpretations 
according to the superhump model. The presence of an extensive number of 
periodicities can explain many peculiarities in the light curve of the 
system that have been previously discussed by many authors.

\begin{table}
\caption{The periods in V1405~Aql}
\begin{tabular}{@{}clll@{}}


Number & Period          & Nature     & Reference \\


1      & 2.8 ms          & spin       & Boirin et al. (2000)   \\
       & 3.7 ms          &            & Galloway et al. (2001) \\
       &                 &            &                        \\
2      & 2979.3(1.1) s   & negative   & this work              \\
       & 2984.6(6.8) s   & superhump  & Smale et al. (1989)    \\
       &                 &            &                        \\
3      & 3000.6508(9) s  & orbital    & Chou et al. (2001)     \\
       & 3000.6(2) s     &            & Homer et al. (2001)    \\
       &                 &            &                        \\
4      & 3027.5510(52) s & positive   & Chou et al. (2001)     \\
       & 3027.555(2) s   & superhump  & Homer et al. (2001)    \\
       &                 &            &                        \\
5      & 3.9087(8) d     & apsidal    & Chou et al. (2001)     \\
       &                 & precession &                        \\
       &                 &            &                        \\
6      & 4.86 d          & nodal      & Chou et al. (2001)     \\
       & 4.74(5) d       & precession & Homer et al. (2001)    \\


\end{tabular}

\end{table}

\subsection{A new relation for superhumps}

Patterson (1999) proposed that the negative superhump deficit is about half 
the positive superhump excess (Section~1.2). As the 3028-s period is about 
0.9 percent longer than the 3001-s period, the corresponding ratio in 
V1405~Aql (0.79) is somewhat larger than this. 
We have checked the connection between the deficit and excess of negative 
and positive superhumps in CVs and found a new relation. Table~2 presents 
the data on the periods in systems showing both types of superhumps.
In Fig.~5 we show this ratio for these systems, and we see a clear 
trend as a function of orbital period. Our result for V1405~Aql fits this 
trend very well. 

\begin{table*}
\begin{minipage}{300 mm}

\caption{Properties of systems that have the two kinds of superhumps}

\begin{tabular}{@{}lccccccc@{}}

Object    & Orbital & Positive  & Positive Superhump      & Negative  & Negative Superhump       & $\phi=\epsilon_{-}/\epsilon_{+}$ & Ref. \\
          & Period  & Superhump & Excess ($\epsilon_{+}$) & Superhump & Deficit ($\epsilon_{-}$) &                             &            \\
          & [d]     & [d]       &                         & [d]       &                          &                             &            \\

AM~CVn&0.011906623(3)&0.0121666(13)& 0.02183(12)         &0.01170613(35)& -0.016839(30)          & -0.7714(57)                 & 1,2,3     \\

V1405~Aql&0.034729754(11)&0.035041099(60)& 0.0089648(20)  &0.034483(13)& -0.0071(7)              & -0.79(8)                    & 4,5       \\

V503~Cyg  &0.0777(2)& 0.08104(7)& 0.0430(27)              &0.07597(18)& -0.022(5)                & -0.51(16)                   & 2,3,6     \\

V1974~Cyg&0.0812585(5)&0.08506(11)& 0.0468(14)            & 0.07911(5)& -0.02644(62)             & -0.565(31)                  & 2,3,7,8   \\

TT~Ari  &0.1375511(2)& 0.1492(1)& 0.0847(7)               & 0.1329(3) & -0.034(3)                & -0.401(40)                  & 2,3,9     \\
 
V603~Aql& 0.1381(1)  & 0.1460(7)& 0.0572(51)              & 0.1343(3) & -0.028(2)                & -0.490(86)                  & 2,3,10    \\

TV~Col  & 0.22860(1) &0.2639(35)& 0.154(15)               & 0.2160(5) & -0.0551(23)              & -0.357(55)                  & 3,11      \\


\end{tabular}
\end{minipage}

$^1$Skillman et al. (1999);
$^2$Patterson (1998);
$^3$Patterson (1999);
$^4$Chou et al. (2001);
$^5$This work;
$^6$Harvey et al. (1995);
$^7$Retter, Leibowitz \& Ofek (1997);
$^8$Skillman et al. (1997);
$^9$Skillman et al. (1998);
$^{10}$Patterson et al. (1997);
$^{11}$Retter et al. 2002b; 

\vspace{0.25cm}
Notes to Table~2:
1. V1159~Ori (Patterson et al. 1995), V592~Cas (Taylor et al. 1998), 
PX~And and BH~Lyn (Patterson 1999) were rejected from this sample as 
their superhump periods are still uncertain (Patterson, personal 
communication).
2. ER~UMa (Gao et al. 1999) was not considered as the detection of 
negative superhumps does not seem secure.

\end{table*}

\begin{figure}

\centerline{\epsfxsize=3.0in\epsfbox{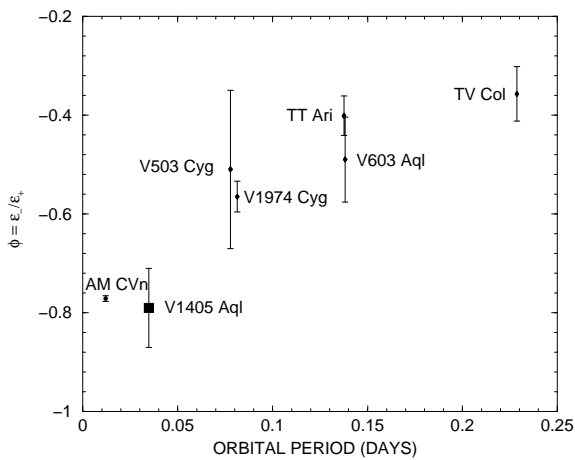}}

\caption{The relation between the orbital period and the ratio between the
negative superhump deficit and the positive superhump excess in systems
that have both types of superhumps.} 

\end{figure}

\subsection{Implications}

Our results strengthen the observational link between positive and negative 
superhumps, and supports the idea that the two types of superhumps have a 
similar physical origin, namely a precessing accretion disc. Retter et al. 
(2002b) speculated that every permanent superhump system may have both 
kinds of superhumps. The data on V1405~Aql support this idea.

Wood, Montgomery \& Simpson (2000) showed that a tilted accretion disc can 
explain the presence of negative superhumps, however, it is still unclear 
what physical force drives the disc to precess in the nodal plane (Murray 
\& Armitage 1998). Wynn (personal communication) suggested that a strong 
magnetic field of the primary white dwarf can cause the disc to precess in 
the nodal plane. If this idea is confirmed, negative superhumps are expected 
in every permanent superhump LMXRB, as their primaries have magnetic fields 
typically much stronger than in CVs.

Montgomery (2001) and Montgomery et al. (in preparation) developed analytic 
expressions for the two types of superhumps. The theoretical values for 
the ratio $\phi$=$\epsilon_{-}/\epsilon_{+}$ mentioned above, were found to 
be consistent with the observations of CVs. Montgomery et al.'s results can 
also explain the relation shown in Fig.~5 for CVs (see their Fig.~6). Using 
their equation 9 \& Fig.~4  (but baring in mind the difference in primary 
mass) and the observed $\phi$ parameter in V1405~Aql (Table~2), we can 
derive a weak upper limit on its mass ratio of $q\la0.08$. For a primary 
neutron star at the Chandrasekhar limit, this implies an upper limit on the 
mass of the secondary -- $M_{2}\la0.12M_{\odot}$, which is presumably a 
helium white dwarf. 

The fact that binary systems with very different configurations (CVs -- a
hydrogen-rich white dwarf and a red companion; AM~CVn -- two degenerate
helium white dwarfs and V1405~Aql -- a neutron star and a helium white 
dwarf) obey the relation presented in Fig.~5 is quite surprising. It might 
be a coincidence, or it may suggest that $\phi$ indeed depends on the 
orbital period and not on the mass ratio. Further determinations of this
parameter in more permanent superhump systems are required to study this issue.


\subsection{The mechanism for negative superhumps}

The mechanism which places the negative superhump signal into the X-ray 
light curve is unclear, but changing vertical disc structure seems a 
promising candidate. It is clear that positive superhumps coincide with 
increased disc thickness (Billington et al. 1996), the evidence being 
consistent with the idea that the area of the disc involved in producing 
the superhump light thickens at the time of the increased dissipation. In 
the nodal precession model, Wood et al. (2000) showed how the passage of the 
secondary star past each of the two halves of the disc out of the midplane 
increases the dissipation in that half. They also required that the side of 
the disc closest to the orbital plane is the most disrupted. If, by analogy 
with positive superhumps, this disruption increases the vertical extent of 
the disc, then it will increase the obscuration of the X-ray source. As the 
disc precesses, the timing of the obscuration will change with phase, 
producing the required modulation at the negative superhump period. The 
lack of structure and large range in phase of the negative superhump 
modulation implies that both the disc structure and occulted region are large, 
the latter presumably being the inner accretion disc or a corona. This 
scenario can also explain why negative superhump signals have not been 
observed so far in the X-ray light curves of low inclination systems.

\subsection{The mechanism for positive superhumps}

Haswell et al. (2001) pointed out that the dissipation of energy in the 
disc, which is believed to be responsible for the positive superhump 
mechanism in CVs in the optical, cannot be applied to LMXRBs. They 
suggested that instead the disc area is changing with the superhump period, 
and thus predicted that superhumps would appear mainly in low inclination 
systems. Figs.~1 \& 2 show that the X-ray data is modulated with the positive 
superhump (f$_{2}$). However, when the dips are rejected from the light 
curve, this peak disappears. Thus in V1405 Aql only the dips show the 
positive superhump, presumably since their amplitude and / or phase vary 
with the apsidal precession period. Note that the phase jitter of the dips 
is modulated with the 3.9-d period (Chou et al. 2001). Therefore, a simple 
explanation of this behaviour is that thickening of the disc rim, which 
causes an increase obscuration of the X-ray source, forms the positive 
superhump. This is consistent with our suggestion for the mechanism for the 
negative superhump in the X-ray (previous section). The disc thickening would 
allow the disc to intercept more of the irradiating flux, thus producing 
optical superhump light seen at all inclinations in a similar way to the 
Haswell et al. area effect. In addition, though, it would introduce dips into 
the X-ray and perhaps optical orbital light curves, allowing the superhump to 
be visible in high inclination systems, in contrast to the prediction of 
Haswell et al. 

\section{Acknowledgments}

We owe a great debt to the anonymous referee for very useful comments that
significantly improved the paper. We thank Michele Montgomery for sending us 
an early version of her paper prior to publication. AR was supported by PPARC 
and is currently supported by the Australian Research Council.


\begin{thebibliography}{99}

\bibitem{b2} Arenas J., Catalan M.S., Augusteijn T., Retter A., 2000, 
MNRAS, 311, 135


\bibitem{b11} Becker R.H., Smith B.W., Swank J.H., Boldt E.A., Holt S.S.,
Serlemitsos P.J., Pravdo S.H., 1977, ApJ, 216, L101

\bibitem{b11} Billington I., Marsh T.R., Horne K., Cheng F.H., Thomas G.,
Bruch A., O'Donoghue D., Eracleous M., 1996, MNRAS, 279, 1274

\bibitem{b2} Boirin L., Barret D., Olive J.F., Bloser P.F., Grindlay J.E.,
2000, A\&A, 361, 121

\bibitem{b2} Callanan P.J., 1993, PASP, 105, 961

\bibitem{b2} Callanan P.J., Grindlay J.E., Cool A. M., 1995, PASJ, 47, 153

\bibitem{b9} Chou Y., Grindlay J.E., Bloser P.F., 2001, ApJ, 549, 1135

\bibitem{b9} Church M.J., Dotani T., Balucinska-Church M., Mitsuda K., 
Takahashi T., Inoue H., Yoshida K., 1997, ApJ, 491, 388

\bibitem{b9} Church M.J., Parmar A.N., Balucinska-Church M., Oosterbroek T.,
dal Fiume D., Orlandini M., 1998, A\&A, 338, 556

\bibitem{b11} Doxsey R.E., Bradt H.V., Dower R.G., Jernigan J.G., Apparao 
K.M.V., 1977, Nat, 269, 112

\bibitem{b11} Galloway D.K., Chakrabarty D., Muno M.P., Savov P., 2001, ApJ,
549, L85

\bibitem{b1} Gao W., Li Z., Wu X., Zhang Z., Li Y., 1999, ApJ, 527, L55

\bibitem{b3} Grindlay J.E., 1986, in: The evolution of galactic X-ray
binaries; Proceedings of the NATO Advanced Research Workshop,
Rottach-Egern, West Germany, June 17-20, 1985 (A87-23176 08-90).
Dordrecht, D. Reidel Publishing Co., p. 25

\bibitem{b3} Grindlay J.E., Bailyn C.D., Cohn H., Lugger P.M., Thorstensen 
J.R., Wegner G., 1988, ApJ, 334, L25

\bibitem{b3} Harvey D.A., Skillman D.R., Patterson J., Ringwald F.A.,
1995, PASP, 107, 551

\bibitem{b3} Haswell C.A., King A.R., Murray J.R., Charles P.A., 2001,
MNRAS, 321, 475

\bibitem{b3} Homer L., Charles P.A., Hakala P., Muhli P., Shih I-C., 
Smale A.P., Ramsay G., 2001, MNRAS, 322, 827

\bibitem{b3} Ko Y.K., Mukai K., Smale A.P., White N.E., 1999, ApJ, 520, 292 

\bibitem{b3} Montgomery M.M., 2001, MNRAS, 325, 761

\bibitem{b3} Morley R., Church M.J., Smale A.P., Balucinska-Church M.,
1999, MNRAS, 302, 593

\bibitem{b3} Murray J.R., 2000, MNRAS, 314, L1

\bibitem{b3} Murray J.R., Armitage P.J., 1998, MNRAS, 300, 561

\bibitem{b4} Nelson L.A., Rappaport S.A., Joss P.C., 1986, ApJ, 304, 231

\bibitem{b11} O'Donoghue D., Charles P.A., 1996, MNRAS, 282, 191

\bibitem{b3} Osaki Y., 1996, PASP, 108, 39

\bibitem{b3} Patterson J., 1998, PASP, 110, 1132

\bibitem{b4} Patterson J., 1999, in "Disk Instabilities in Close Binary 
Systems", eds. Mineshige S., Wheeler C., Universal Academy Press, 61


\bibitem{b3} Patterson J., Jablonski F., Koen C., O'Donoghue D., Skillman 
D.R., 1995, PASP, 107, 1183

\bibitem{b3} Patterson J., Kemp J., Saad J., Skillman D., Harvey D., Fried
R., Thorstensen J.R., Ashley R., 1997, PASP, 109, 468

\bibitem{b4} Priedhorsky W.C., Terrell J., 1984, ApJ, 280, 661


\bibitem{b4} Retter A., Naylor T., 2000, MNRAS, 319, 510

\bibitem{b4} Retter A., Leibowitz E.M., Ofek E.O., 1997, MNRAS, 286, 745

\bibitem{b4} Retter A., Chou Y., Bedding T., 2002a, ASP Conf. Ser., Vol ???, 
eds. B.T. Gaensicke, K. Beuermann, K. Reinsch., in press 

\bibitem{b4} Retter A., Hellier C., Augusteijn T., Naylor T., Bembrick C., 
McCormick J., Velthuis F., 2002b, MNRAS, submitted

\bibitem{b6} Scargle J.D., 1982, ApJ, 263, 835

\bibitem{b3} Schmidtke P.C., 1988, AJ, 95, 1528

\bibitem{b3} Skillman D.R., Harvey D., Patterson J., Vanmunster T., 1997,
PASP, 109, 114

\bibitem{b3} Skillman D.R. et al., 1998, ApJ, 503, L67

\bibitem{b3} Skillman D.R., Patterson J., Kemp J., Harvey D.A., 
Fried R., Retter A., Lipkin Y., Vanmunster T., 1999, PASP, 111, 1281

\bibitem{b3} Smale A.P., Lochner J.C., 1992, ApJ, 395, 582

\bibitem{b3} Smale A.P., Mason K.O., White N.E., Gottwald M., 1988, MNRAS, 
232, 647

\bibitem{b3} Smale A.P., Mason K.O., Williams O.R., Watson M.G., 1989, 
PASJ, 41, 607

\bibitem{b3} Smale A.P., Mukai K., Williams O.R., Jones M.H., Corbet R.H.D.,
1992, ApJ, 400, 330


\bibitem{b3} Taylor C.J. et al., 1998, PASP, 110, 1148

\bibitem{b3} Walter F.M., Bowyer S., Mason K.O., Clarke J.T., Henry J.P., 
Halpern J., Grindlay J.E., 1982, ApJ, 253, L67

\bibitem{b3} Warner B., 1995, Cataclysmic Variable Stars, Cambridge 
University Press

\bibitem{b3} White N.E., 1989, A\&A Rev., 1, 85

\bibitem{b9} White N.E., Swank J.H., 1982, ApJ, 253, L61

\bibitem{b4} Whitehurst R., 1988, MNRAS, 232, 35                 

\bibitem{b4} Whitehurst R., King A., 1991, MNRAS, 249, 25

\bibitem{b3} Wood M.A., Montgomery M.M., Simpson J.C., 2000, ApJ, 535, L39


\bibitem{b3} Yoshida K., Inoue H., Mitsuda K., Dotani T., Makino F.,
1995, PASJ, 47, 141

\end{thebibliography}
\end{document}